\documentclass[journal=jacs,manuscript=article,layout=onecolumn]{achemso}

\usepackage[version=3]{mhchem} 
\usepackage{graphicx}
\usepackage{amsmath}
\usepackage{amssymb}
\usepackage{amsfonts}
\usepackage{soul}
\usepackage{dcolumn}
\usepackage{bm}
\usepackage{hyperref}
\usepackage[mathlines]{lineno}
\usepackage{xcolor}
\usepackage{titlecaps}
\definecolor{sph}{rgb}{0.0588, 0.3216, 0.7294} 
\definecolor{ppk}{rgb}{1.0, 0.4549, 0.0902} 
\newcommand{\nc}{\newcommand}
\nc{\nn}{\nonumber}
\nc{\txt}{\textrm}
\nc{\txtsup}{\textsuperscript}
\nc{\txtsub}{\textsubscript}
\nc{\calL}{\mathcal{L}}
\nc{\U}{\mathcal{U}}
\nc{\T}{\mathcal{T}}
\nc{\E}{\mathcal{E}}
\nc{\calH}{\mathcal{H}}
\nc{\sect}[1]{{\sl #1 .--}}
\nc{\xx}[1]{\textcolor{black}{#1}}
\nc{\XX}[1]{\textcolor{black}{#1}}
\nc{\YY}[1]{\textcolor{red}{#1}}

\newcommand{\onlinecite}[1]{\hspace{-1 ex} \nocite{#1}\citenum{#1}} 

\author{Subhajit Sarkar}
\email{sbhjt72@gmail.com , subhajis@srmist.edu.in}
\affiliation{Department of Physics and Nanotechnology,
SRM Institute of Science and Technology
Kattankulathur-603 203, India.}
\affiliation{Institute of Theoretical Physics, Jagiellonian University, Lojasiewicza 11, 30-348, Krakow, Poland,}

\author{Yonatan Dubi}
\affiliation{Department of Chemistry, Ben-Gurion University of the Negev, Beer Sheva 84105, Israel,
}
\affiliation{Ilse Katz Center for Nanoscale Science and Technology, Ben-Gurion University of the Negev, Beer Sheva 84105, Israel.
}

\date{\today}

\title{Time Crystals from single-molecule magnet arrays}

\makeatletter
\newcommand*{\forcekeywords}{
  \acs@keywords@print
  \let\acs@keywords@print\relax
}

\begin{document}
\forcekeywords


\begin{center}
\date{\today}
\end{center}

\begin{abstract}
Time crystals, a unique non-equilibrium quantum phenomenon with promising applications in current quantum technologies, mark a significant advance in quantum mechanics. Although traditionally studied in atom-cavity and optical lattice systems, pursuing alternative nanoscale platforms for time crystals is crucial. Here we theoretically predict discrete time-crystals in a periodically driven molecular magnet array, modeled by a spin-S Heisenberg Hamiltonian with significant quadratic anisotropy, taken with realistic and experimentally relevant physical parameters. Surprisingly, we find that the time-crystal response frequency correlates with the energy levels of the individual magnets and is essentially independent of the exchange coupling. The latter is unexpectedly manifested through a pulse-like oscillation in the magnetization envelope, signaling a many-body response. These results show that molecular magnets can be a rich platform for studying time-crystalline behavior and possibly other out-of-equilibrium quantum many-body dynamics.
\end{abstract}

\maketitle
Time crystals (TC) are a genuine non-equilibrium phase of an interacting quantum system that breaks the time translation invariance, despite the Schr\"odinger equation being time-translation invariant \cite{Zaletel_RMP, Wilczek_prl_classical_tc, Wilczek_prl_quantum_tc, Else_review_2020, Sacha_2017,yao2020classical, gambetta2019classical,Sacha_PRA_DTC}. Discrete Time-Crystals (DTCs), in particular, have generated significant interest due to their distinctive property of spontaneously breaking discrete time-translation symmetry \cite{Zaletel_RMP, beatrez2023critical, peng2021floquet}. DTCs are periodically driven systems where local observables display indefinite, robust, and coherent oscillations at an integer multiple of the driving period across a broad spectrum of initial states~\cite{Zaletel_RMP}. DTCs are proposed to aid quantum technologies by increasing the stability and sensitivity of nuclear magnetic resonance (NMR) spectroscopy, and as protocols for quantum-enhanced metrology \cite{Zaletel_RMP, cabot_param_sense_PRL_2023}. Furthermore, their use across various physical platforms enables benchmarking and validation of noisy intermediate-scale quantum devices \cite{mi2021time, Preskill2018quantumcomputingin, zhang2023characterizing}.

Multiple strategies exist to achieve time-crystalline order in closed and open systems, encompassing localization, prethermalization, dissipation, and error correction \cite{Zaletel_RMP}. Numerous experiments, including predominantly optical experiments involving atom-cavity systems, spin ensembles, quantum processors \cite{Choi2017, Zhang2017, Kyprianidis1192, pal_prl_cluster, magnon_space_time_crystals_prl, Dissipative_DTC_PhysRevLett_atom_cavity, smits_prl_superfluid_QG, Zhang2017, taheri2022all, Dogra_science.aaw4465, Autti2021, mi2021time,  beatrez2023critical, peng2021floquet}, and solid-state semiconductor-based quantum dots (QD) arrays \cite{Qiao2021, vandyke2020protecting, randall2021many, Economu_PhysRevB_2019}, have been shown to exhibit DTC behavior. 

The interplay between the interacting degrees of freedom, driving, and dissipation in quantum systems can stabilize the oscillatory dynamics in a very long time limit, leading to the formation of continuous TCs in undriven systems and DTCs in periodically driven systems \cite{RieraCampeny2020timecrystallinityin, diss_Flq_DTC_PRL_Ikeda, BTC_PhysRevLett_Fazio, carollo2021exact, Gong_PhysRevLett, Marino_Demler_NJP_2019, Gambetta_PhysRevLett_metaDTC, Prosen_SciPostPhys_2020, Tindall_2020, chinzei2021criticality, Gunawardana2021DynamicalLI, sarkar2022protecting, buca2021algebraic, Buca_prb_2020, xiang2023selforganized, Buca_PRX, Buca_nat_comm2019, Buca_PhysRevLettQG_2019, sarkar2021signatures, Sarkar_2022gex, alaeian2022exact}. Beyond the scientific interest in DTCs as a fundamental non-equilibrium phenomenon, suggestions for utilizing DTCs in future applications have already been put forward, for example, as a simulator of complex quantum networks \cite{estarellas2020simulating}, characterizing quantum processors \cite{zhang2023characterizing}, and as quantum sensors and quantum thermodynamic engines \cite{montenegro2023quantum, cabot_param_sense_PRL_2023, paulino2022nonequilibrium}.
 
 The recent interest in this phenomenon naturally leads to the following question: Going beyond the atom-cavity and optical lattice systems, what other nanoscale platforms can exhibit DTC over sufficiently long and experimentally measurable time scales? As a partial answer to this question, we have recently demonstrated that a quantum dot (QD) array placed between electronic leads in a paradigmatic experimental setting can identify and directly measure the appearance of DTC \XX{{\sl via}} oscillations in the transport current \cite{Sarkar_2022gex, sarkar2021signatures}.
 \begin{figure}
\includegraphics[scale=0.34]{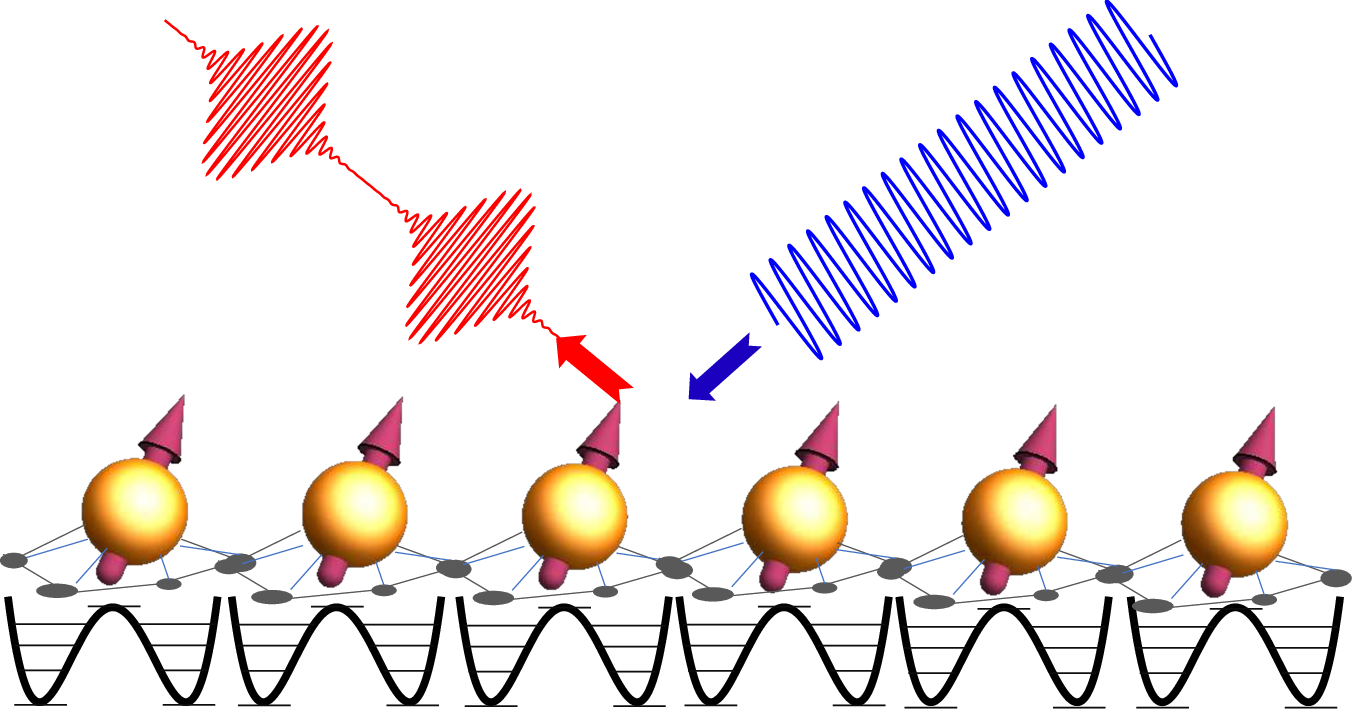}
\caption{\textbf{Schematic of the set-up:} A Heisenberg exchange-coupled chain of single molecular magnets (SMMs) is depicted, where each SMM is represented by an arrow-ball, with a double well illustrating its characteristic energy-level diagram. A spin-$S$ SMM exhibits $(2S+1)$ states, with $m_s = \pm S, \pm (S-1), \dots, \pm1$ doubly degenerate states, and a non-degenerate state for $m_s = 0$. The underlying discrete time crystal (DTC) behavior converts the blue continuous wave into a red subharmonic pulse train. The gray molecular structures in the background represent the molecular environments.
}
\label{fig0_schematic}
\end{figure}

In this study, we demonstrate that Single-Molecule Magnets (SMMs) showcase nontrivial time-crystalline behavior, marking a significant shift from the commonly studied optical setups. SMMs are metallo--organic molecules that exhibit a finite magnetic moment at the single-molecule level \cite{friedman2010single, zabala2021single}. They have drawn considerable attention for nearly three decades due to their promising possibilities of storing and processing quantized information at the molecular level \cite{moreno2021measuring, lockyer2022five, bode2023dipolar}, and promising candidates as quantum simulators \cite{chicco2023proof}, embedded quantum error correction \cite{chiesa2021embedded}. For example, a one-dimensional column of collinear magnetic moments of [TbNcPc]$^+$ promotes an intermolecular ferromagnetic coupling with the necessary properties for quantum information processing \cite{katoh2018control}. Despite these promising efforts, molecular qudits-based quantum simulators have faced challenges due to the lack of experimental implementations \cite{carretta2021perspective}, and the full range of phenomena that single-molecule magnets can exhibit remains largely unexplored. Here we show theoretically that SMMs, either isolated or coupled \XX{{\sl via}} nearest-neighbor exchange interactions, can exhibit a DTC with realistic physical parameters. 
  
We consider a ferromagnetic exchange-coupled chain of SMMs with $S \ge 1$ (see schematic description in Fig.~\ref{fig0_schematic}) and show that the system exhibits a stable subharmonic oscillation corresponding to the DTC state, that originates from the energy levels of individual molecular magnets and a pulse profile that corresponds to the many-body effect. Having no disorder, our DTC can be classified as a clean DTC \cite{Pizzi_clean_2020}, compared to the usual disorder-induced DTC in closed systems \cite{Else_PRL_Floquet_DTC, Khemani_PRL_DTC, khemani2019brief, Dziarmaga_PhysRevB_2DTC}. \XX{Additionally, we show that, due to the distinctive energy landscape of SMMs, the SMM arrays behave as a ``wave converter", generating a pulse train out of continuous-wave (CW) driving.} Moreover, given that DTCs have been experimentally observed in several other quantum-computing platforms, \XX{\textit{e.g.}}, superconducting qubits \cite{mi2021time}, trapped-ion-based simulator \cite{Kyprianidis1192, randall2021many}, semiconductor-based spin-qubits \cite{Qiao2021}, our theoretical proposal can serve as a benchmarking tool for recently proposed quantum simulator based on SMM-qudits \cite{chicco2023proof, chiesa2021embedded}.

\section{Results and Discussion} 

\textbf{The System:} We consider a periodically driven chain of $N$ ferromagnetically coupled SMMs, represented by the anisotropic Hamiltonian \cite{zabala2021single}
\begin{equation}\label{eqn:Hamiltonian}
    \calH = - J~\sum_{j=1}^{N-1} \mathbf{S}_{j} \cdot \mathbf{S}_{j+1} - \sum_{j=1}^{N} \big[ D (S_{j}^{z})^{2} + E (S_{j}^{x})^{2} - E(S_{j}^{y})^{2} \big] + \mathbf{B}(t)\cdot \sum_{j=1}^{N} \mathbf{S}_j,
\end{equation}
where $S_{j}^{x}, S_{j}^{y} $ and $S_{j}^{z}$ are the three components of the spin
operators corresponding to the $j'$th SMM. The first term is the exchange term, with the exchange coupling energy $J$. The term $\displaystyle \left[ D (S_{j}^{z})^{2} + E (S_{j}^{x})^{2} - E(S_{j}^{y})^{2} \right]$ corresponds to the anisotropic Hamiltonian of an individual SMM, where $D>0$ and $E>0$ are the quadratic and rhombic anisotropies, respectively, that defines the energy landscape of the SMM. These anisotropy parameters are determined, \XX{\textit{e.g.}}, \XX{{\sl via} the} high-frequency electron paramagnetic resonance or neutron spectroscopy \cite{barra1996superparamagnetic, Caciuffo_PhysRevLett}. For a typical SMM, e.g., $\text{Mn}_{12}-\text{acetate}$, the system parameters are, $D /k_{B} = 0.56~K$ and $E/k_{B} = 4.5\times10^{-3}~K$, indicating the single-ion anisotropy being the dominant energy scale, and we can neglect the rhombic anisotropy term for most of the purposes (cf. \onlinecite{Petiziol2021}). We consider here chains of up to $N=5$, in line with synthesized SMM arrays which were limited to a few ($\le 6$) SMM \cite{katoh2018control}.

The last term $ \mathbf{B}(t)\cdot\mathbf{S}_{j}$ represents `Zeeman energy' corresponding to the externally applied magnetic field. We consider a circularly polarized periodic drive along with a constant longitudinal magnetic field in the $z-$ direction,
\begin{equation}
     \mathbf{B}(t)\cdot\sum_{j=1}^{N}\mathbf{S}_{j} =  \sum_{j=1}^{N} \left[ \frac{B}{2} \left(S_{j}^{+} e^{-i \omega t} + S_{j}^{-} e^{i \omega t} \right) + B^{\prime} S_{j}^{z}\right],
\end{equation}
where \(B = g \mu_{B} B_{ext} \) (and similarly, \(B' = g \mu_{B} B^{\prime}_{ext} \)) with $B_{ext}$ being the external magnetic field, \XX{\textit{e.g.}}, expressed in `Gauss' in CGS unit or `Tesla(T)' in SI unit, $\mu_B$ being the Bohr magneton and $g$ being the Lande-g factor. Moreover, circular polarization allows for a static Hamiltonian in the Floquet (rotated) frame, crucial for no net energy absorption from the drive and thus preventing DTC heating. However, linearly polarized drives, common in magnetic resonance and SMM experiments, rely on the Rotating Wave Approximation. This permits controlled energy exchanges vital for spin manipulation but detrimental to DTC's long-lived sub-harmonic oscillations.

From the theoretical point of view, the model we consider goes beyond the existing prethermal-DTC physics of spin$-S$ (non-disordered) systems corresponding to a binary Hamiltonian involving the alternate Ising model and transverse field \cite{Pizzi_classical_DTC}. In contrast to impurities in diamonds and trapped ions \cite{Zhang2017, Choi2017}, our SMM-inspired model can be more thermalization prone due to various anisotropies. 

To investigate the possibility of coherent DTC oscillations in the system, we calculate the spin-dynamics in general and dynamics of local magnetization $S_{j}^{z}$ in particular, in an SMM array (Eq.~\ref{eqn:Hamiltonian}) starting from a thermal state using the Sch\"{o}dinger equation for the density matrix $\rho$, $ \frac{d \rho}{dt} = -i [\calH, \rho]~$ (we set 
$\hbar = 1$), and evaluate the dynamics of local magnetization, $\langle S^{z}_{j} \rangle = \text{Tr}[S^{z}_{j} \rho(t)]$. In particular, we take the initial density matrix to be a thermal density matrix, $\displaystyle \rho = \frac{e^{-\beta \calH}}{\text{Tr}\left[ e^{-\beta \calH} \right]}$ at an inverse temperature $\beta = (JS)^{-1}$ for $S = 1$ and we keep $\beta = \mathcal{O}[(J)^{-1}]$ for $S>1$. As we discuss in the following sections, whether the temperature is low or high is essentially determined by the gap between the ground and the first excited states. For our system, this gap is determined by the amplitude of the external drive and is $\sim \mathcal{O}(B + |D|)$.

\begin{figure}
\includegraphics[scale=0.35]{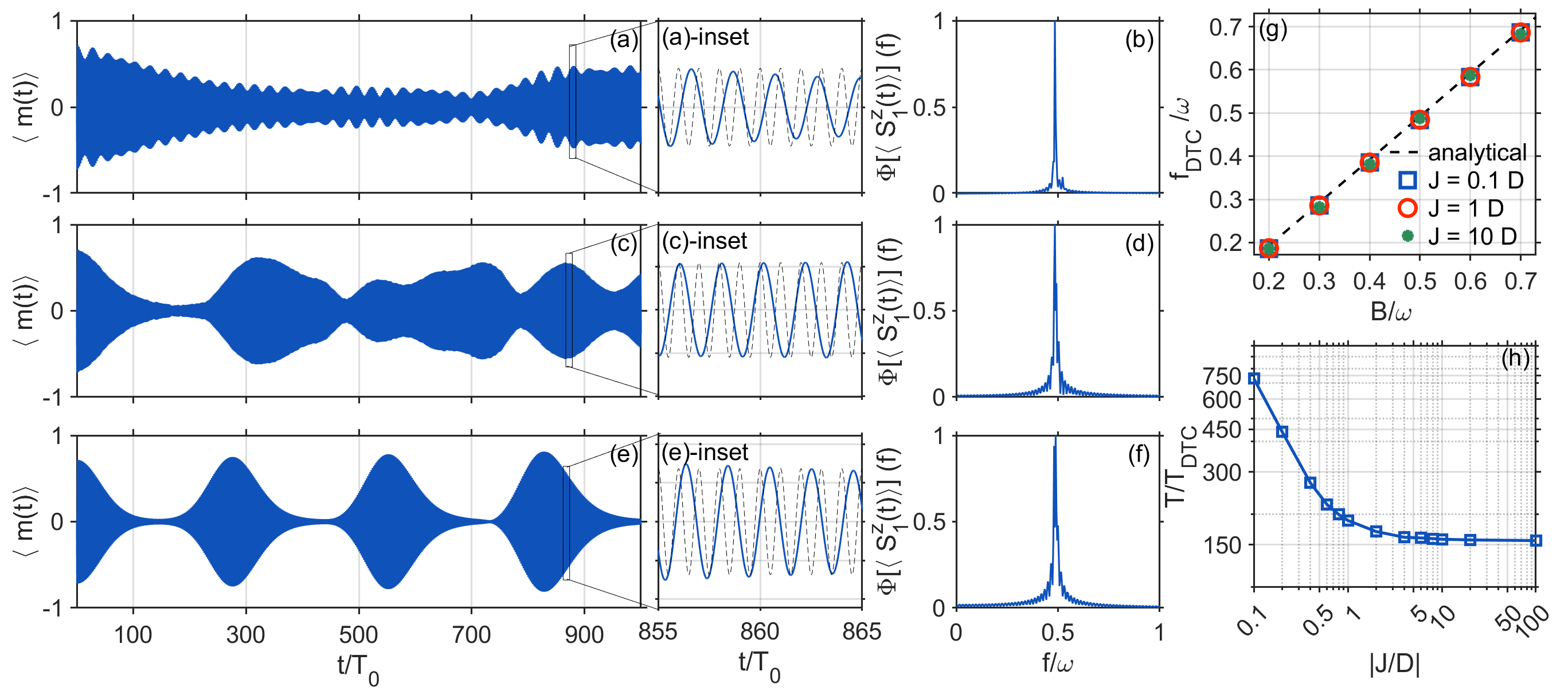}
\caption{Plots of average magnetization $\langle m(t) \rangle$ as a function of $t/T_0$ up to 1000 [(a),(c),(e)] correspond to $J=0.1 D$, $J= D$, and $J=10 D$, respectively, for a chain of $N=5$ exchange-coupled SMMs. [(a)-inset,(c)-inset,(e)-inset] plot $\langle m(t) \rangle$ as a function of $t/T_0$, zoomed-in between $855T_0$ and $865T_0$. (b) (d) (f) is the discrete Fourier transform of the magnetization, viz., $\Phi \left[ \langle S^z_j (t)\rangle \right](f)$. The sub-harmonic frequency $f_{DTC} = 0.49~\omega$; (g) the sub-harmonic frequency $f_{DTC}/\omega$ as a function of $B/\omega$, the amplitude of the external oscillatory field {for different values of exchange coupling J}. The solid line corresponds to the analytical expression for the DTC frequency; (h) the ratio of envelope periodicity (determined from the time duration between two successive minima in the envelop oscillation amplitude in SI Fig. 3) to the DTC periodicity, $T/T_{\text{DTC}}$ as a function of $|J/D|$ plotted on a log-log scale {for $B = \omega/2 $ and $N=3$}.
}
\label{fig1_osc_DFT}
\end{figure}

 \textbf{Magnetization dynamics:} We start by showing that, in a suitable parameter regime, the system exhibits subharmonic oscillations in local magnetization $\langle S^{z}_{j}(t) \rangle$, namely a DTC state. As a specific example, we first consider an individual SMM that has $S=1$. The other parameters of the system include the anisotropy parameter $D = 0.206~ \mu$eV or $~50$ MHz, the driving frequency, $\omega = 5 \times (2\pi D) = 1.57~ \text{Grad/s}$, and the period $T_0 = \frac{2\pi}{\omega} = \frac{1}{5D}\sim 4~ \text{ns}$. We further consider that the static Zeeman field $B'$ is set so that $B' - \omega = 0$, ensuring that the mean magnetization remains zero. The amplitude of the periodic magnetic field varies within $0.1 \omega <B<\omega$, translating to an external magnetic field $B_{ext} = \frac{\hbar \omega}{g \mu_{B}} = 8.93~mT = 89.3$ Gauss, for $g\approx 2$ and $\mu_B = 57.88~\mu eV /T$, as the upper limit of the external magnetic field. Despite our choice of parameters falling within the scope of the present experimental capabilities \cite{Luis_Mettes_Jongh_ch5, Collett_PhysRevResearch}, we normalize all parameters relative to $D$ to make our results universally applicable to any SMM array. We choose the exchange coupling as a parameter within the range $0.1D < J < 10D$. This is motivated by the recent implementation of quantum gates and quantum simulators \cite{chiesa_molecular_2024} using SMMs achieved through the modular design of molecular qubits with controllable exchange coupling in the regime $J>D$ \cite{ferrando2016modular}.

Fig.\ref{fig1_osc_DFT} illustrates the long-time dynamics of average magnetization $\displaystyle \langle m(t) \rangle$ for different values of exchange coupling. Fig.\ref{fig1_osc_DFT}(a), (c), and (e) plot the average magnetization $\displaystyle \langle m(t) \rangle = \frac{1}{N} \sum_{j=1}^{N} \langle S^{z}_{j}(t) \rangle $ as a function of time up to 1000 driving periods for three different values of exchange coupling, namely, $J=0.1 D$, $J= D$, and $J=10 D$, respectively. The discrete Fourier transform (DFT)  $\Phi \left[ \langle S^{z}_{j}(t)\rangle \right](f)$ corresponding to the above three different values of $J$ are plotted in Fig. \ref{fig1_osc_DFT}(b), (d), and (f), respectively, showing the (approximate) period doubling corresponding to a DTC response with DTC frequency, $f_{DTC} \approx \frac{\omega}{2}$. Here $f$ represents the frequency variable for the DFT, which is distinct from the driving frequency $\omega$. 

\textbf{Sub-harmonic frequency:} To evaluate the dependence of the sub-harmonic frequency $f_{\text{DTC}}$ on the amplitude of the external driving field $B$, $\displaystyle \frac{f_{\text{DTC}}}{\omega}$ as a function of $\displaystyle \frac{B}{\omega}$ is plotted in Fig. \ref{fig1_osc_DFT}(g), for three values of exchange coupling, $J=0.1D$ (squares),$J=1D$ (circles),$J=10D$ (dots). Surprisingly, even though $J$ is comparable to other energy scales of the system, the DTC frequency remains essentially independent of $J$. This is further illustrated by the analytical form of the sub-harmonic frequency $\displaystyle f_{\text{DTC}} = \sqrt{B^2 + \left(\frac{D}{2}\right)^2} - \frac{D}{2}$ for an array of $S=1$ SMMs, shown in Fig.~\ref{fig1_osc_DFT}(g) as a black dashed line. This indicates $\displaystyle f_{\text{DTC}}$ is primarily determined by the energy levels of a single molecular magnet, as we discuss below in detail. This result underscores the fundamental role of single-molecule properties in the observed time-crystalline behavior.

\textbf{The role of exchange coupling:} Although $J$ (which represents the many-body interaction in the system) does not seem to affect the DTC frequency, it significantly impacts the system's dynamics. Specifically, the envelope of the oscillations depends on $J$ (Fig.~\ref{fig1_osc_DFT}(a, c, e). Fig.~\ref{fig1_osc_DFT}(h) shows the dependence of the period of the {\sl envelope oscillation} of the magnetization on $J$ for $N=3$. This indicates that (i) with weak coupling between the SMMs the period of the envelop oscillation is several hundred times larger than the DTC period, and (ii) the envelope period saturates with increasing values of $J$. This indicates that the envelope dynamics are influenced by many-body interactions. In the weak exchange coupling case, \XX{\textit{e.g.}}, corresponding to $J = 0.1D$, Fig.\ref{fig1_osc_DFT}(a) shows another period that is a few tens of the DTC period, and the corresponding envelope oscillation exhibits a decaying trend in the long time limit. This is evidenced by comparing envelope amplitude around $100T_0$ and $500T_0$. Thus, although DFT in Fig.\ref{fig1_osc_DFT}(b) exhibits a tiny side-peak, it will eventually merge with the sub-harmonic peak leading to a divergent envelope period in the chain of non-interacting SMMs. 

\begin{figure}
\includegraphics[scale=0.3]{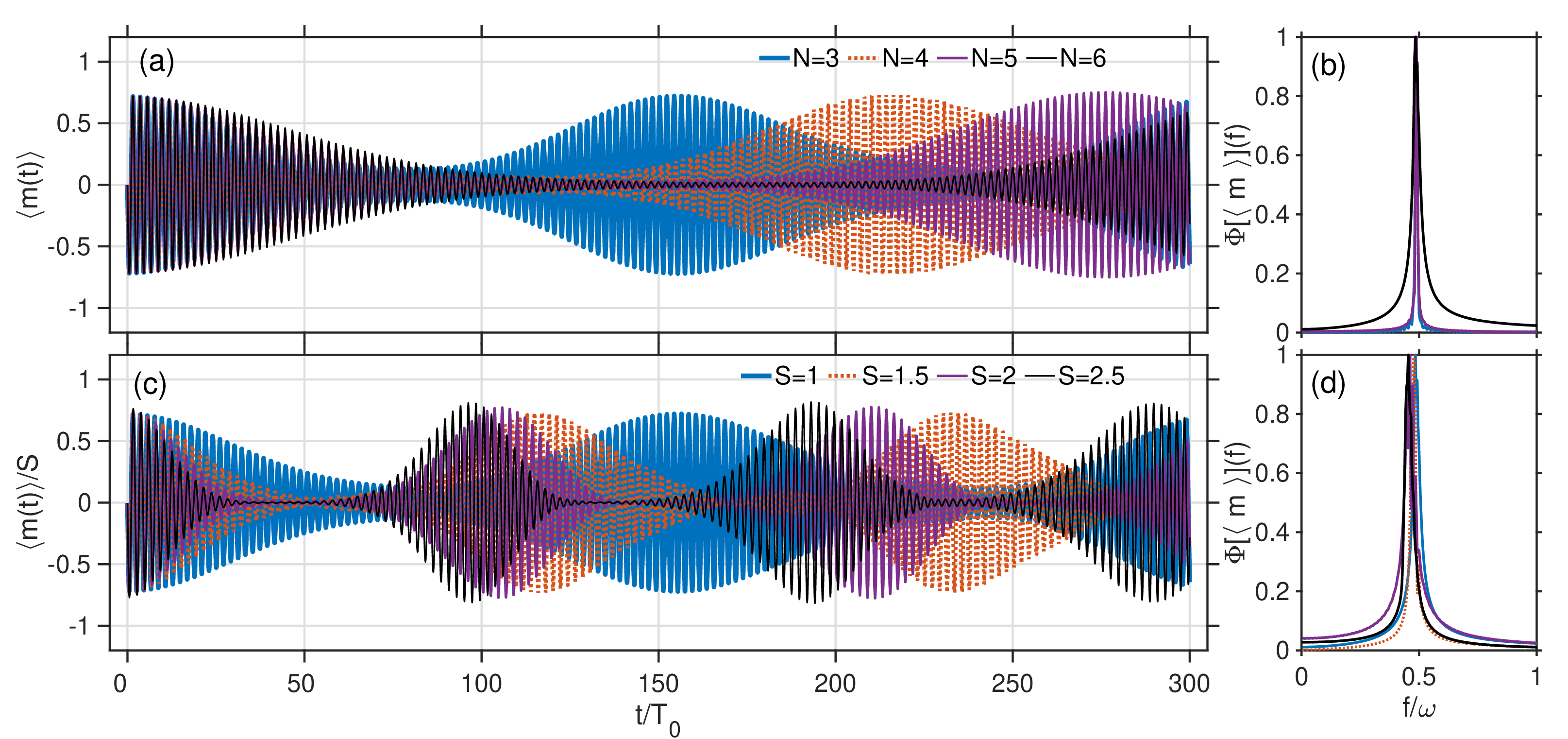}
\caption{Plots of (a) average magnetization $\langle m(t)\rangle$ as a function of normalized time $t/T$ for different values chain length $N = 3 \text{ to } 6$ for $J = 10D$ for $S=1$; (b) the corresponding DFTs, (c) average magnetization $\frac{\langle m(t)\rangle}{S}$ (i.e., normalized with respect to $S$ to make the magnitudes comparable) as a function of normalized time $t/T$ for different values spin $S = 1,1.5,2,2.5$ for $J = 10D$ for $N=3$, and (d) the corresponding DFTs.
}
\label{fig2_osc_Ndep}
\end{figure}

The saturation of the envelope period signifies that the pulse-like profile of $\displaystyle \langle m(t) \rangle$ in Fig.\ref{fig1_osc_DFT}(e) is a true many-body effect and would remain robust against the order of magnitude change in $J$. We further emphasize that the envelope period is also contingent on the chain length. Fig.~\ref{fig1_osc_DFT}(h) corresponds to $N=3$, and for a different chain length, the value of the envelope period would be higher, as can be seen in the naked eye from Fig.~\ref{fig1_osc_DFT} (a, c, e) and further corroborated in Figure \ref{fig2_osc_Ndep}.

Figure \ref{fig2_osc_Ndep} provides a detailed analysis of the impact of chain length and spin value on the time evolution of average magnetization $\langle m(t) \rangle$ under strong exchange interactions. Figure \ref{fig2_osc_Ndep}(a) shows the time evolution of $\langle m(t) \rangle$ for varying chain lengths ($N = 3$ to $6$) with $S = 1$ under $J = 10D$. It reveals that the envelope periods of the oscillations increase with chain length, suggesting that in the Large$-N $ limit, the oscillations will eventually decay to zero for a fixed spin value $S$. Figure \ref{fig2_osc_Ndep}(b) confirms the sub-harmonic oscillations despite reducing amplitude, as illustrated by the discrete Fourier transform (DFT) of the corresponding oscillations.

Figure \ref{fig2_osc_Ndep}(c) shows $\langle m(t) \rangle$, normalized with respect to $S$, against time for different spin values ($S = 1, 3/2, 2, 5/2$) with a fixed chain length $N = 3$. In contrast to Figure \ref{fig2_osc_Ndep}(a), It demonstrates that higher spin values result in decreased envelope periods, indicating enhanced robustness of the DTC oscillations.  Figure \ref{fig2_osc_Ndep}(d) corroborates this by showing that the sub-harmonic oscillation frequency remains consistent across different spin values, as seen in the DFT.

Figure \ref{fig2_osc_Ndep} highlights two opposing trends: increasing chain length $N$ leads to longer envelope periods and eventual decay of oscillations, whereas increasing spin values $S$ enhances the stability and persistence of DTC oscillations by reducing the envelop period. Although for a fixed spin $S$, increasing $N$ suggests that DTC would decay in time, supporting Figure 5(c) shows that the time-scale of the decay, defined as the time over which the amplitude of the oscillation decays to $1/e$ of its peak value, increases roughly linearly with $N$ with a hint of saturation at very large $N$. This indicates longer chains also stabilize DTC oscillations for a longer duration, before an eventual decay. Therefore, achieving indefinite stabilization of DTC requires both large $N$ and $S$, and for any finite and large $S$ the DTC behavior is akin to a prethermal DTC \cite{Kyprianidis1192} in systems with $S \geq 1$.

To further elucidate the interplay of exchange coupling and energy levels of the individual SMMs, we consider different initial density matrices, both synchronized and non-synchronized ones, see Supporting Information Sec. II. By a synchronized initial state we refer to a state for which all sites in the system have identical local magnetizations. With weak interaction and non-synchronized initial states, we observe oscillations reflective of the spin dynamics of individual SMMs that further evolve to a noisy blend with stronger exchange interactions. Synchronized states yield a unified DTC frequency, irrespective of interaction strength, see Supporting Fig. 3. Intriguingly, even non-synchronized states eventually synchronize across all sites, leading to quantum synchronization, albeit at the expense of DTC.

\textbf{Analytical form of frequency:} To understand how the energy levels of a single molecular nano-magnet determine the sub-harmonic response we make a unitary transformation using $U = \prod_{j}e^{-i S_{j}^{z} \frac{\omega t}{2}}$ to a rotated frame, such that the time-periodic Hamiltonian becomes a static external field, viz., $\displaystyle \calH_{\text{F, ext}}= (B, 0, B^{\prime} - \omega) \cdot \sum_{j} \mathbf{S}_{j}$. In this case, one can choose the external longitudinal magnetic field in the $z-$ direction to cancel with the frequency, viz., $B^{\prime} - \omega = 0$. Note that this choice is independent of the value of spin $S$ of the SMM. Therefore, the effective Hamiltonian in the rotating frame is, $\displaystyle \calH_{F} = \sum_{j} \left[- J~ \mathbf{S}_{j} \cdot \mathbf{S}_{j+1} - D (S_{j}^{z})^{2} + B S_{j}^{x} \right]$, corresponding to a transverse field Heisenberg model with quadratic anisotropy, (see Supporting Information section III for calculational details). Therefore the energy levels of a single molecular nano-magnet are given by the effective Hamiltonian $\displaystyle \calH_{F,j}^{\text{SMM}} =  -D (S_{j}^{z})^{2} +B S_{j}^{x} $.

For the SMMs with $S=1$, the eigenvalues of $\displaystyle \calH_{F,j}^{SMM}$ are given by $\E_{g}= \left(-\sqrt{B^2+(D/2)^2}-\frac{D}{2}\right) $ for the ground state and $\E_{e,1} = -D$, and $\E_{e,2} =\left(\sqrt{B^2+(D/2)^2}-\frac{D}{2}\right)$ for the excited states, respectively. The energy levels that participate in the sub-harmonic generation are the ground state ($|g_{j}\rangle$ for the $j'$th SMM) $\E_g$ and the first excited state ($|e_{1,j}\rangle$ for the $j'$th SMM) $\E_{e,1}$, leading to the frequency, $\E_{e,1} - \E_g = \displaystyle f_{\text{DTC}} = \sqrt{B^2 + \left(D/2\right)^2} - \frac{D}{2} $. This form is plotted in the solid line in Fig.~\ref{fig1_osc_DFT}(g). 

\begin{figure}[h]
\includegraphics[scale=0.16]
{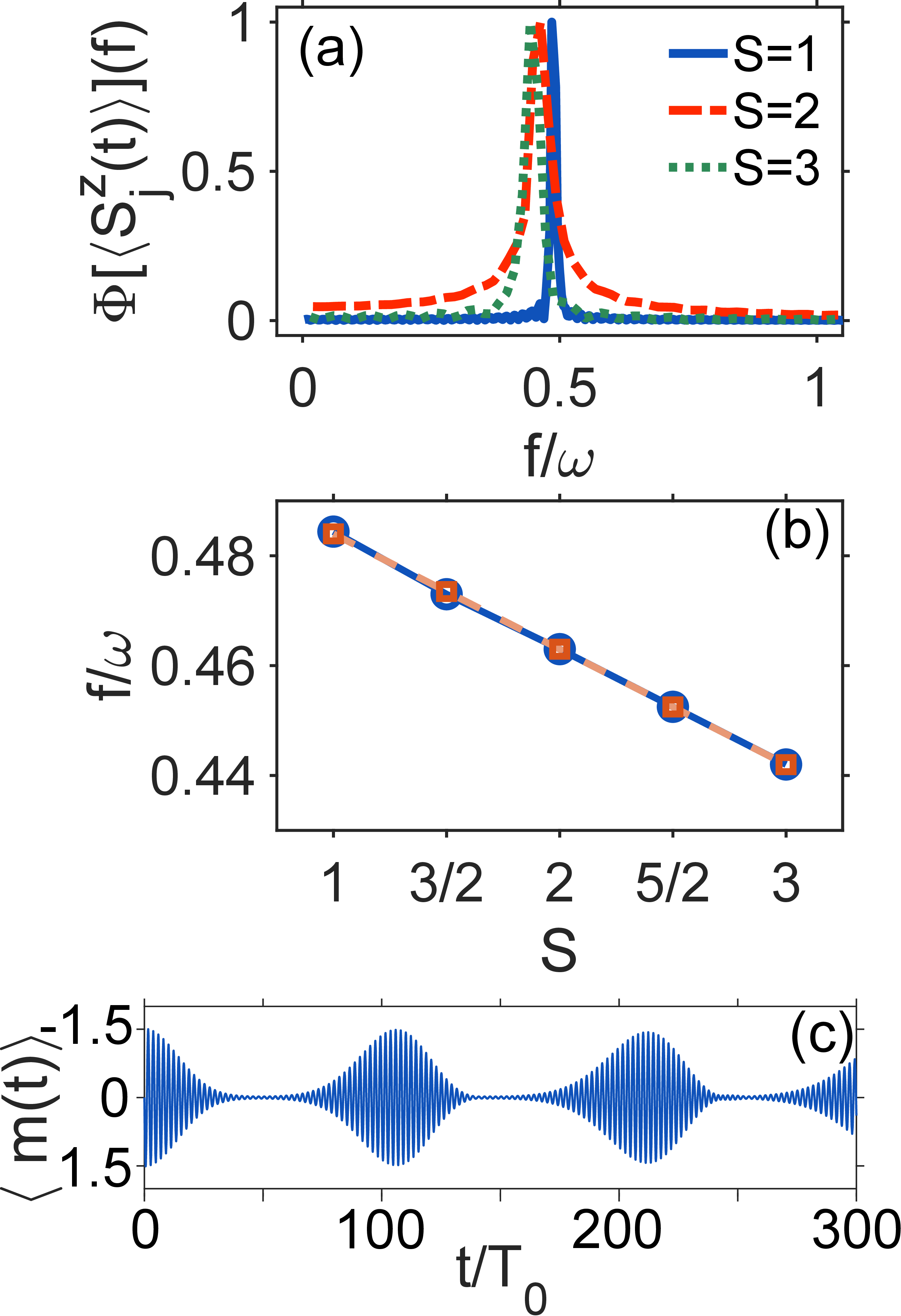}
\caption{Plots of (a) discrete Fourier transform $\Phi \left[ \langle S^z_j (t)\rangle \right](f)$ of the on-site magnetization as a function of frequency $f/\omega$ for different spin values; (b) the sub-harmonic frequency $f_{DTC}/\omega$ as a function of the value of spin-$S$, the orange dashed line corresponds to the value of $\E_{e,1} - \E_g$. (c) A plot of average magnetization $\langle m(t) \rangle$ as a function of $t/T_0$ up to 300 driving periods corresponding to a chain of 3 SMMs with $S=2$.
}
\label{fig2_high_spin_DTC_freq}
\end{figure}

\textbf{Higher spin cases:} These results for the $S=1$ system indicate that, in general, for a chain of SMMs with any value $S$, the ground and the first excited states of the individual SMM would take part in the DTC oscillation. To verify this, in Fig. \ref{fig2_high_spin_DTC_freq} (a) we plot the DFT $\Phi[S_{j}^{z}(t)](f)$ of on-site magnetization $S_{j}^{z}(t)$ as a function of $f/\omega$ for $S=1, 2,~\text{and}~3$. Fig. \ref{fig2_high_spin_DTC_freq} (a) shows that sub-harmonic oscillation frequency remains similar in order of magnitude. This further verifies that, for all $S$, the sub-harmonic oscillation frequency is primarily determined by the energy levels of the individual SMMs. Fig. \ref{fig2_high_spin_DTC_freq} (b) further shows the DTC frequencies against the spin values for $S =1,~ 3/2,~2,~5/2,~3$, and they match with $\E_{e,1} - \E_g$.

For higher spin values the profile of $\langle m(t) \rangle$ develops the shape of a pulse train. Fig. \ref{fig2_high_spin_DTC_freq} (c) plots the average magnetization $\langle m(t) \rangle$ as a function of time for over 300 driving periods showing such a pulse train. The amplitude of $\langle m(t) \rangle$ remains undamped indefinitely, showcasing the stability of the DTC oscillation. This shows that a chain of exchange coupled SMMs can convert the continuous wave drive of frequency $\omega$ into a pulse train of sub-harmonic frequency $f<\omega$. This remains true for larger chain lengths. For example, the combination $S=1$ and $N=5$ brings in the pulsating character of the DTC oscillation in Fig.\ref{fig1_osc_DFT}(e). Comparing Fig.\ref{fig1_osc_DFT}(e) and Fig. \ref{fig2_high_spin_DTC_freq} (c) we can conclude that any moderately finite (and therefore, experimentally achievable) chain of coupled SMM with $S\geq 1$ would indeed convert a continuous wave drive of frequency $\omega$ into a pulse train of sub-harmonic frequency $f<\omega$. Such conversion of continuous wave (CW) to pulse-train in the context of time crystals or for periodically driven $S=\frac{1}{2}$ Heisenberg chains is, to the best of our knowledge, a unique phenomenon of SMMs.



\textbf{Conditions for sub-harmonic generation:} The DTC oscillations shown in Figs. \ref{fig1_osc_DFT} and \ref{fig2_high_spin_DTC_freq} (c) correspond to the condition that {$B' - \omega = 0$}. This further leads to the condition that if the amplitude of the periodic drive $B<\omega$ the average magnetization oscillates with a sub-harmonic frequency, and the DTC is seen in $\langle m(t) \rangle$. In particular, we have considered $B=0.5\omega$. In this case, the $x-$ and $y-$ components of the spins oscillate with a higher-harmonic frequency to maintain the conservation of spin, i.e., $\displaystyle \sum_{\alpha=(x,y,z)}\left[S_{j}^{\alpha} (t)\right]^2 = S(S+1)$, see \textit{Supporting Fig. 4}, and Ref. \onlinecite{sarkar2021signatures, Sarkar_2022gex}. Conversely, if $B>\omega$ the average magnetization oscillates with a higher-harmonic frequency, however, the corresponding $x-$ and $y-$ show sub-harmonic response. In this case, the DTC is observable in the $x-$ and $y-$ components of the spins (see Supporting Information section IV).

\textbf{Mechanism of time-crystal:} Ref. \onlinecite{Zaletel_RMP} clearly defines the DTC as a long-lived oscillation of any local observable, such as the magnetization in our case, at a period that is a multiple of the driving period. Therefore, to substantiate our identification of sub-harmonic oscillations as signatures of a DTC, we examine three key criteria: {(i) indefinite (long-time) persistence, signifying stability, (ii) robustness of oscillation frequency against system parameter perturbations, and (iii) originating from interacting many-body effects.}

The long-time persistence of these oscillations is evidenced by their non-diminishing amplitude over time, excluding the overall envelope modulation attributable to exchange coupling, as illustrated in Fig.\ref{fig1_osc_DFT}(a), (c), and (e). This modulation introduces an additional oscillation period in $\displaystyle \langle m(t) \rangle$, complementing the fundamental DTC period, thereby affirming criterion (i).

The robustness of the oscillation frequency (criterion (ii))  is validated by Figs. \ref{fig1_osc_DFT}(b), (d), and (f) which demonstrate that despite an order of magnitude change in the value of $J$ the sub-harmonic frequency remains the same. Hence, the oscillation in $\displaystyle \langle m(t) \rangle$ meets all stipulated conditions to be classified as a DTC \cite{Pizzi_clean_2020}. Additionally, the susceptibility of the DTC, defined as \(\displaystyle \chi_{\text{DTC}} = \lim_{B \rightarrow 0} \frac{\partial f_{\text{DTC}}}{\partial B} \rightarrow 1\) when \( B \gg D\), indicates that the DTC frequency is robust against perturbations in the driving. Therefore, the condition \( B \gg D\) is a key condition for a robust sub-harmonic response.

We now consider condition (iii). In our system, the many-body effects come from the exchange coupling $J$. When $J \rightarrow 0$, the oscillation frequency represents the Rabi frequency of the collection of non-interacting SMMs. Quite surprisingly, in our system, even for $J \neq 0$, the DTC frequency $f_{\text{DTC}}$ remains fixed at the Rabi frequency of an isolated SMM, which might lead one to conclude that the oscillation in the coupled system is a `many-body Rabi oscillation' (and not a TC state).

However, the situation is more nuanced. For an Ising coupling, \( \displaystyle \sum_{j=1}^{N} J S_{j}^{z} S_{j+1}^{z}\), the oscillations do not survive in the coupled system (see Supporting Information sec IV) indicating the importance of the nature of the coupling. The $SU(2)$ symmetry of the isotropic Heisenberg exchange coupling plays a significant role in the emergence of a `many-body Rabi' oscillation. An external periodic field generates the largest gap -- between the ground state and the first excited state -- in the many-body spectrum, primarily determined by the magnetic field \(B\) with minor corrections from the anisotropy parameter \(D\) (note $B \gg D$). 

Put simply, despite the DTC frequency seeming like the Rabi frequency of individual SMM, the states participating in DTC are the many-body states, viz., the ground state and the first excited state, the coherent superposition of all possible single spin-flip states. This validates condition (iii). In the rotated (Floquet) frame, the oscillation represents a Larmor precision. Such a Larmor precision was categorized as a `prethermal' continuous time crystal in an undriven system in the context of a spin \(\frac{1}{2}\)--XY model \cite{khemani2019brief}. This, along with the stability of the oscillation for an indefinitely long time and robustness against any perturbative change of system parameters, motivates us to conclude the `many-body Rabi oscillation' as a DTC.

The underlying mechanism of DTC is the appearance of a dynamical symmetry operator \cite{Buca_PRX, Buca_nat_comm2019, Buca_PhysRevLettQG_2019, Buca_prb_2020, sarkar2021signatures, sarkar2022protecting, Sarkar_2022gex, alaeian2022exact}. Although an exact dynamical symmetry operator could not be identified, we show in the Supporting Information that the total spin raising operator is $\tilde{S}^{+}_{tot} = \sum_{j} \left( S_{j}^{z} + i S_{j}^{y} \right)$, exhibiting overlap with the magnetization, is an approximate dynamical symmetry, and $\mathfrak{Re}(\tilde{S}^{+}_{j})$ shows DTC oscillation. For the spin operator \(\tilde{S}^{+}_{\text{tot}}\) to act as a dynamical symmetry, the following conditions must be met: the external field \(B\) must be much stronger than the anisotropy \(D\), the spin \(S\) of the system should be large, and only low energy excitations (low temperatures) should be present. As we show in the Supporting Information, under these conditions the dynamical symmetry eigenvalue is given by, \(\left(B -\frac{D}{2} (S - n/N) + \frac{(S+1)D^2}{16B}\right) + \mathcal{O}(\frac{D^3}{B^2})\), where $n$ is the total number of spin-flips from the ground state (magnons) of the system. Quite importantly, the necessity of a small number of spin-flips (in other words, low-lying magnons) for the appearance and stability of our DTC further corroborates the many-body nature of DTC corresponding to condition (iii). A single spin-flip (1-magnon excitation), although it appears to represent a non-interacting system, actually reflects many-body effects, as magnons are collective excitations originating from the nearest-neighbor spin interactions \cite{ashcroft_mermin_1976}.

In Fig. \ref{fig2_high_spin_DTC_freq}(b) we show that for all $S$, the sub-harmonic oscillation frequency is primarily determined by the energy levels of the individual SMMs. This conforms with conditions for the dynamical symmetry because, given $B \gg D$, the energy gap between the ground state and the first excited states in both the entire system and the individual SMM is primarily determined by $B$.

\section{ Conclusions} To summarize, we have found that a DTC can be induced by applying a circularly polarized {electromagnetic(EM)-wave} to the exchange-coupled Heisenberg chain of SMMs. We consider an experimentally realizable setup with a relevant parameter regime and numerically demonstrate using the evolution of the mixed-state density matrix. 

Apart from the existence of DTC, we also show that for higher spin values our set-up exhibits an ability to convert a continuous wave drive to a pulse train of sub-harmonic frequency when the amplitude of the external field is less than the frequency of the drive, i.e., $B<\omega$. Conversely, by making $B > \omega$, our setup can act as a tunable higher-harmonic generator.

Our findings establish that the DTC oscillation frequency hinges predominantly on the energy levels of individual SMMs, while the exchange coupling impacts the frequency of the DTC oscillation envelope. One key question remains: what is the origin of the pulsating character of the oscillation envelope? While possible clues lie in the detailed structure of the time-crystal states and how they develop with \XX{\textit{e.g.}}, chain length, at this point we make the following conjecture: the envelope oscillations develop due to the finite velocity of the spin-wave excitations which increases with increasing $J$ and decreases with increasing $S$. This is evident as increasing $J$ and $S$ results in increasing and decreasing the period of the envelope, respectively. 

Additionally, the fate of this DTC hinges on the interplay between dynamical-symmetry-based mechanism due to its origin from the individual SMM energy levels \cite{khemani2019brief, Buca_nat_comm2019, sarkar2021signatures, sarkar2022protecting} and mean-field-based mechanisms due to its stability derived from the large -- $S$ and large-$N$ limits. This is apparent from the conditions -- large--$S$ and low temperatures, therefore, only low-lying excitations -- for the existence of the dynamical symmetry. This interplay determines the specific nature of environmental interactions that the DTC can withstand. The DTC we obtained is fundamentally different from the pre-thermal DTCs typically reported in the literature, which are driven by very high frequencies. \cite{frey2022realization, Russomanno_PRB_2017}. Our DTC corresponds to a pre-thermal DTC based on the Floquet Dynamical Symmetry.

Moreover, the ground and the first excited states remain gapped due to the external magnetic field \cite{chauhan_tunable_2020}. This gap prevents the system from becoming a continuous spectrum in the large -- $S$ limit.

Temperature plays a crucial role in the stability and lifetime of DTC oscillations, as any temperature $k_B T \gtrapprox B$ would allow a substantial population of the higher magnon excited states, thereby melting the DTC. For the parameters we used, this temperature is $\sim 36$ mK, which is achievable in a dilution refrigerator \cite{Thiele_science_2014, ZU2022103390}.

We anticipate that our findings will drive further exploration into the relationship between SMM energy levels, exchange interactions, and DTC pulse-train dynamics. This could potentially reveal links to (semi-) classical pre-thermal phases of matter \cite{Pizzi_classical_DTC}. Thus, our results highlight the SMM array as a key nano-scale system for pioneering out-of-equilibrium experiments.


\section{Methods}
\subsection*{Density Matrix Evolution}

To investigate the time evolution of the system, we compute the dynamics of the density matrix \(\rho(t)\) using the Liouville-von Neumann equation with $\hbar = 1$:
\[
\frac{d\rho(t)}{dt} = -i [H(t), \rho(t)]
\]
where \(H(t)\) is the time-dependent Hamiltonian given in \eqref{eqn:Hamiltonian}. The time evolution of \(\rho(t)\) is carried out numerically using a discretized time step. We take the steps to be $ \Delta t = T_0 / 1000$

\textbf{Initialization} The system is initialized in a thermal state at a given temperature, where the initial density matrix \(\rho(0)\) is given by:
\[
\rho(0) = \frac{e^{-\beta H}}{\text{Tr}(e^{-\beta H})}.
\]

\subsection*{Observables}
The time-dependent expectation value of a local observable, such as the magnetization \(\langle S_z(t) \rangle\), is calculated from the evolving density matrix \(\rho(t)\):
\[
\langle S_z(t) \rangle = \text{Tr}(S_z \rho(t))
\]
where \(S_z\) is the spin operator corresponding to the observable of interest. This quantity is tracked over the entire evolution of the system over time.

To identify periodic behavior and detect sub-harmonic oscillations, the discrete Fourier transform (DFT) of the time-dependent magnetization is computed.
 \section{Acknowledgements} SS acknowledges useful discussions with Michal Rams and Marek Rams, and partial funding from the National Science Center, Poland, under Project 2020/38/E/ST3/00150. SS also acknowledges Srijani Mallik for insightful discussions on dilution refrigerators and for suggesting Ref. \onlinecite{ZU2022103390}.

 \section{Supporting Information}
\titlecap{System size dependence of the DTC frequency, Choice of the initial states, Static Hamiltonian in the rotated frame, Mechanism of the DTC and difference from the Rabi oscillations, Condition for sub-harmonic response}


 


\bibliography{mm}

\newpage

\begin{figure*}
    \centering
   \includegraphics[keepaspectratio=true,scale=0.15]{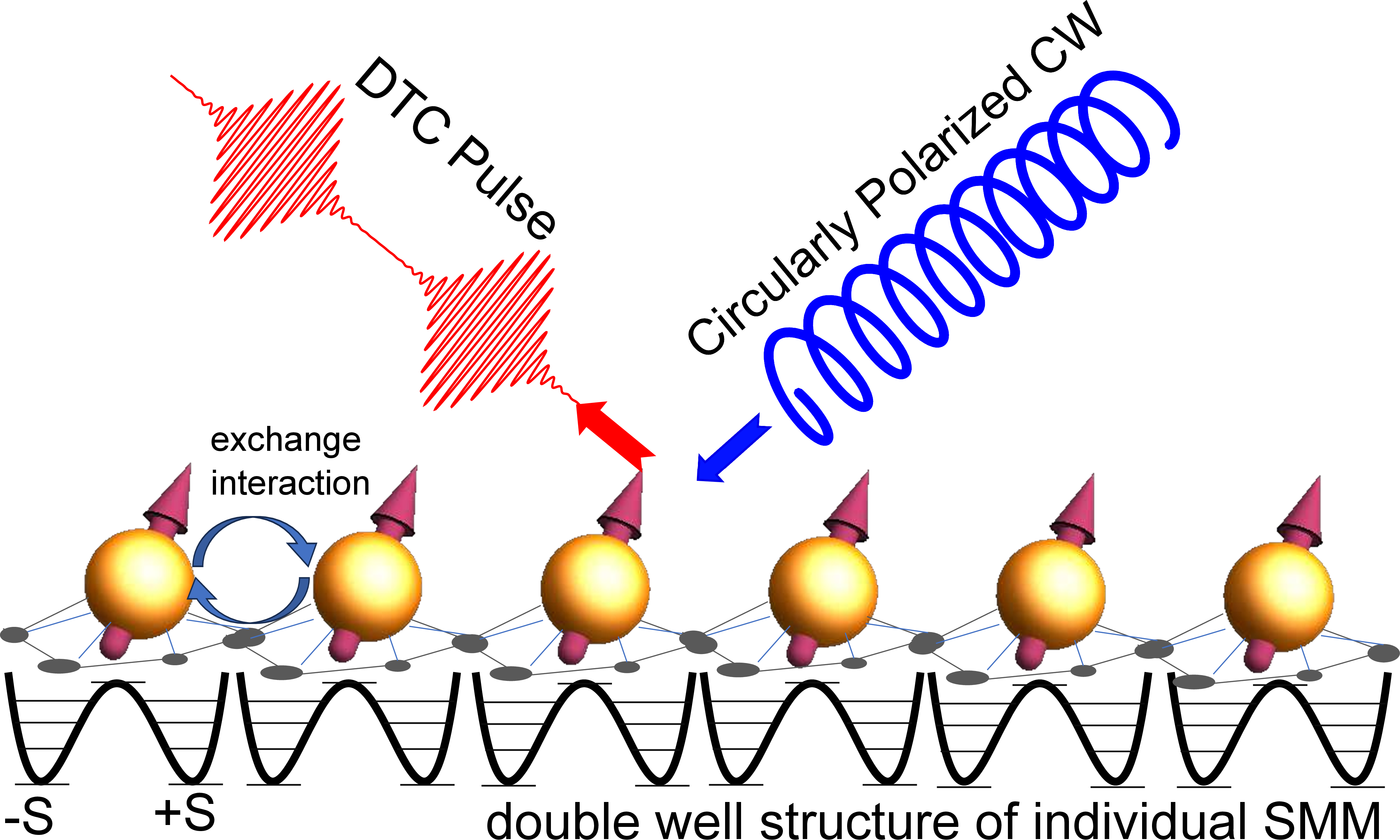}
    \caption*{“For Table of Contents Only”}
    \label{For Table of Contents Only}
\end{figure*}

\newpage

\end{document}